# How to help university students to manage their interruptions and improve their attention and time management

Aurora Vizcaíno, Ignacio García-Rodríguez de Guzmán, Antonio Manjavacas,

Félix García, José A. Cruz-Lemus, Manuel Ángel Serrano

Alarcos Research Group, Universidad de Castilla-La Mancha, Paseo de la Universidad 4, Ciudad Real, 13071, Castilla-La Mancha, Spain

{Aurora.Vizcaino, Ignacio.GRodriguez, Antonio.Manjavacas, Felix.Garcia, JoseAntonio.Cruz, Manuel.Serrano}@uclm.es

## ABSTRACT

Technology has changed both our way of life and the way in which we learn. Students now attend lectures with laptops and mobile phones, and this situation is accentuated in the case of students on Computer Science degrees, since they require their computers in order to participate in both theoretical and practical lessons. Problems, however, arise when the students' social networks are opened on their computers and they receive notifications that interrupt their work. We set up a workshop regarding time, thoughts and attention management with the objective of teaching our students techniques that would allow them to manage interruptions, concentrate better and definitively make better use of their time. Those who took part in the workshop were then evaluated to discover its effects. The results obtained are quite optimistic and are described in this paper with the objective of encouraging other universities to perform similar initiatives.

**Keywords**: higher education; time management; negative effect of interruptions; self-regulated learning; self-efficacy

## INTRODUCTION

We live in a technological world surrounded by constant interruptions originating from multiple internal and external sources. The number of daily interruptions to which we are exposed has risen significantly, owing to the increase in new technologies in our daily lives, something that could impair our performance when doing our jobs or any other type of task in which concentration is essential. The use of technology in our work environments, along with the consequent media multitasking, leads to a situation in which our brains tend to get distracted by interruptions, thus preventing us from keeping focused and staying in the present moment (Was et al., 2019). This is critical in higher education where students attend lectures with their mobile phones and computers on the tops of their desks, meaning that they get distracted by almost all of the notifications that they receive.

Moreover, it has been proved that information and communication technologies (ICTs) contribute to increasing stress levels and to a reduction in time and space awareness, while people and, of course students, tend to fragment the attention paid to what they do (Elliott-Dorans, 2018; Patterson & Patterson, 2017). Concerned about this situation, we decided to study the types of interruptions in students' lives and designed a workshop with the aim of helping them manage the different types of interruptions to which they are exposed and improve their time management. We would like to clarify that not only external interruptions were considered in our study; we also wished to go further and consider the distractions that originate from our own thoughts, which some authors call mind wondering (Was et al., 2019), since attempting to prevent and resolve these types of interruptions has become a challenge as we are usually not aware of their existence. In order to evaluate how useful the workshop was, a study was conducted, and is described in this paper.

The paper is structured as follows: Section 2 shows the motivation and the research method used in this study, while Section 3 describes the consequences of interruptions. In Section 4, a pilot study performed in order to research the effect of training students to manage their interruptions and thoughts is described.

Section 5 describes the study conducted and the results obtained. Finally, our conclusions, contributions to the field and future work are presented in Section 6.

# MOTIVATION AND RESEARCH METHOD

This section provides a description of the research method employed. This work came into being because, after teaching the first year of Computers Science degrees for almost twenty years, we realised that each year students have a shorter attention span, concentrate less and retain less information in their memories. Our concern about this situation led us to observe what was now different with respect to the previous years when we started to teach. **The first phase of this work, therefore, consisted of "observing"** how our students behave in lectures (see *Figure 1*). We realised that all the students have their mobile phones on the tops of their desks and many of them have their laptops open. The students' social networks are open on both devices and they, therefore, receive notifications that interrupt their work and concentration.

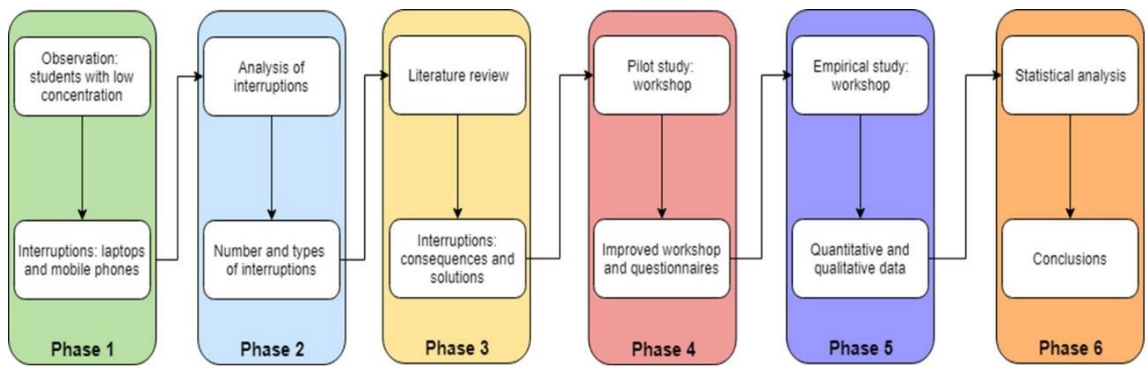

**Fig. 1. Research Method**

The **second phase was to "analyse the situation",** in order to evaluate the scope of what we observed. A questionnaire (see *Appendix A*) was, therefore, prepared in order to discover the number and the type of interruptions to which students are exposed when they are studying and when they are in lectures. 75 second-year students on a Computer Science degree completed the questionnaire. The results obtained showed that the number of interruptions during one study hour is mostly between 0 and 22 (see *Figure 2*), with the majority of students experiencing 6 to 11 interruptions (30.67%). Taking into account that recovering the focus of attention after an interruption takes an average of 10 minutes (Kessler, 2011; Patterson & Patterson, 2017), we can determine that with approximately 10 interruptions, the student's study time may be useless.

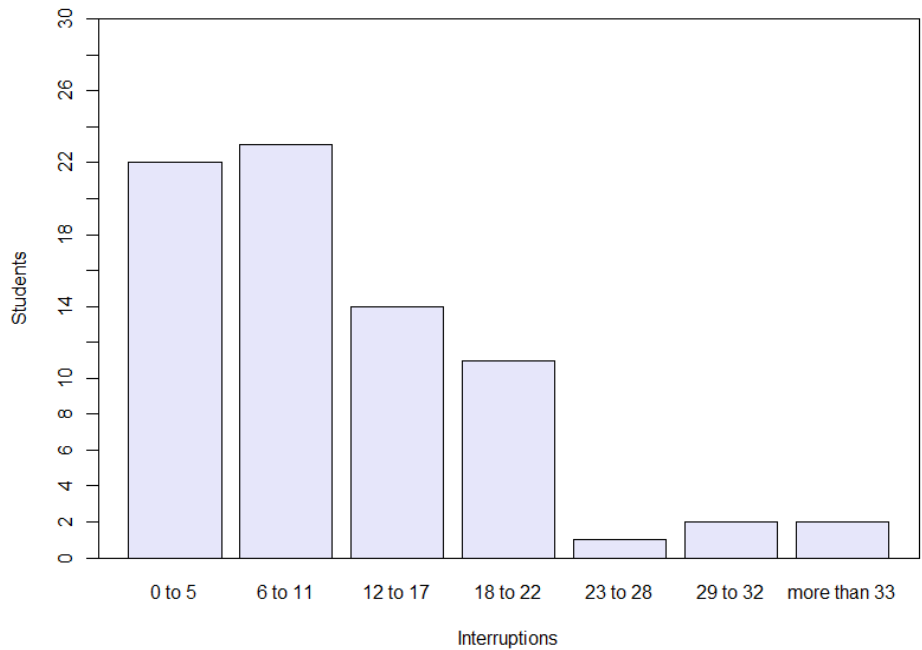

**Fig. 2. Interruptions during one study hour.**

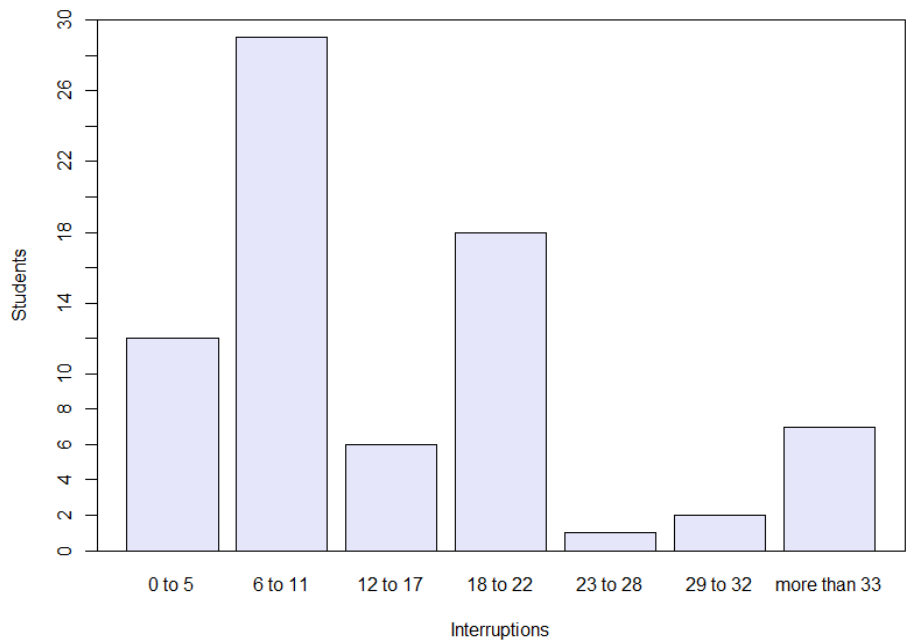

**Fig. 3. Interruptions during one lecture hour.**

What is more, the number of interruptions during one lecture hour (*Figure 3*) ranges principally between 6 and 11 (38.67%), with a considerable number of students having between 18 and 22 interruptions (24%), a significant number that we should not overlook. In this case, the number of interruptions is higher than in one study hour and we can, therefore, assume that more sources of interruptions are present when they are in lectures than when they are studying, which is quite surprising.

With regard to the type of interruptions, we distinguished between external interruptions (*Figure 4*) and internal interruptions (*Figure 5*) and obtained the following results: in the case of external interruptions, the greatest source of interruptions is WhatsApp, followed by the other social media and by people (the label "people" refers to the situation in which one person interrupts another by starting a conversation).

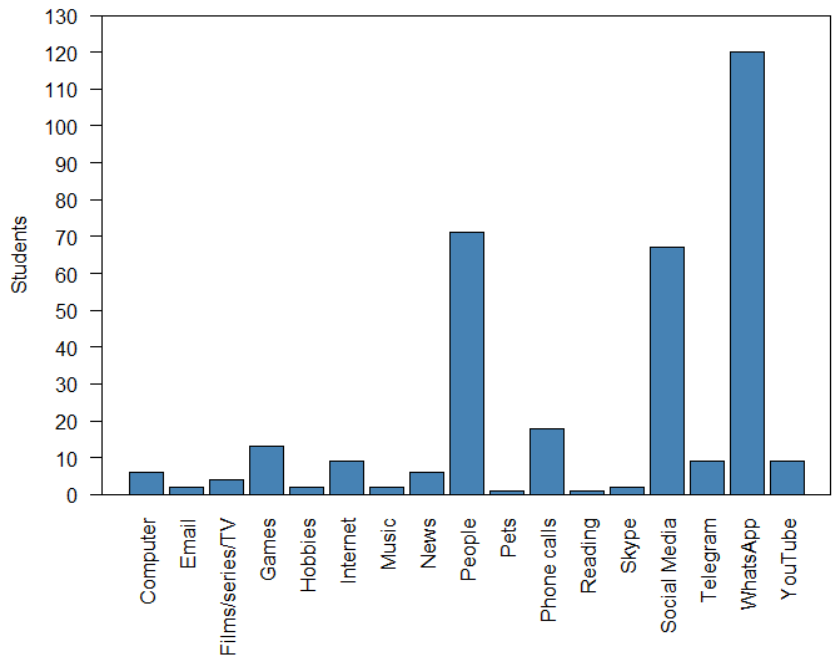

**Fig. 4. Principal external sources of interruption.**

From the viewpoint of internal distractions shown in *Figure 5*, we did not pay too much attention to physiological needs, such as eating, drinking or going to the bathroom, as these tend not to be completely dependent on the individual. We wish mainly to analyse those sources of interruption that the individual really can avoid, as is the case of thoughts, people (see definition above) or fatigue.

The results of Phase 1 and Phase 2 (see *Figure 1*) were the motivation for our work, since we considered that the number of interruptions is very high and that something should be done in order to improve this situation.

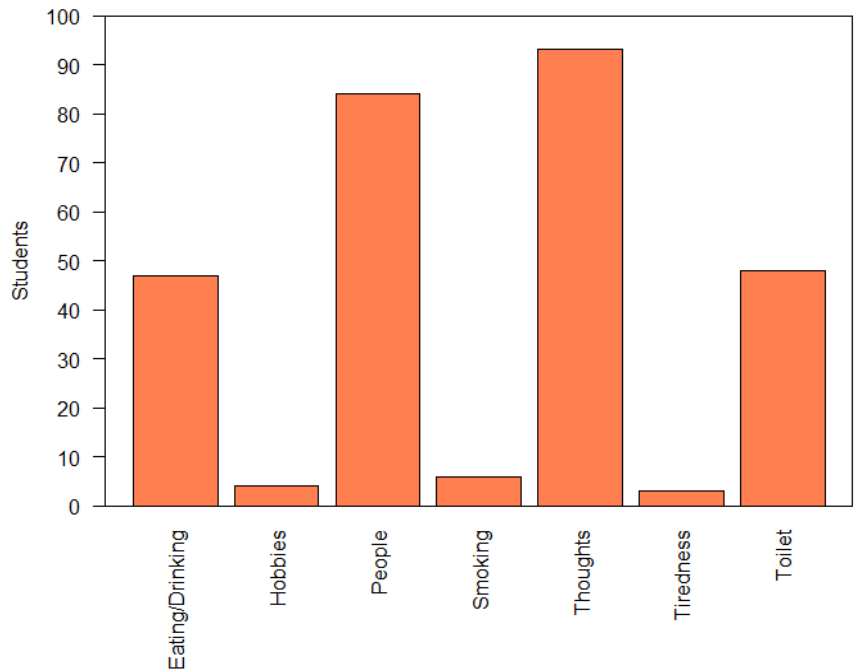

**Fig. 5. Main internal sources of interruption.**

**The Third Phase** was focused on studying literature in order to discover the consequences of interruptions when people perform tasks, to categorize them and tackle how they could be decreased by using suitable techniques, such as mindfulness or time management.

As a result of the findings obtained in the previous phases, a pilot study was conducted in the **Fourth Phase** in order to evaluate whether it is feasible to improve students' attention and concentration by means of training provided in a workshop. This allowed us to evaluate the learning activity and to improve the design and the materials to be used in the empirical study carried out in **Phase 5.** The main research question to be answered as a result of the empirical study concerned how suitable training can improve students' time organization, thought management and levels of concentration. The results were analysed in **Phase 6**.

A more detailed description of Phases 3, 4, 5 and 6 are presented in the following sections, respectively.

# THE EFFECTS OF INTERRUPTIONS ACCORDING TO LITERATURE

It has been empirically demonstrated that interruptions are a vital factor as regards evaluating a subject's task performance and response time when performing a particular activity (Mark et al., 2008). One of the main measures usually taken to reduce their number is to make use of technological tools to solve organizational and communicative problems and the delays related to them. However, as stated in (Levine et al., 2007), the impact of ICTs is neither pre-ordained nor predictable; the counterproductive side effects of using these technologies mean that some users relegate certain interruptions, and pass them on to another user, who is then also interrupted. This means that not only have both users been interrupted, but what actually prompted the interruption is temporarily displaced (Kraushaar & Novak, 2010; Kuznekoff & Titsworth, 2013).

The use of new technologies has simultaneously increased our tendency towards *multitasking*, which causes constant changes in our focus of attention, thus worsening the performance of tasks and increasing the time required to complete them (Iqbal & Horvitz, 2007; Karnowski & Jandura, 2014; Mark et al., 2008). We must consider that the greater the number of changes in our focus of attention, the more prone to errors we will be in our task development (Mark et al., 2008). Furthermore, interruptions lead to several problems as regards subjects' recovering the state they were in prior to the interruption; they also increase the levels of stress, frustration, pressure and effort, as interruptions move us away from the focus of our work (Elliott-Dorans, 2018).

The trend towards using different means of technology encourage young people to simultaneously engage in multiple activities (Altmann et al., 2014). According to Altmann et al. (2014), Yildirim & Dark (2014), and Karnowski & Jandura (2018), multitasking increases distractibility and interferes with learning. This signifies that not only does our productivity not increase, but it actually becomes worse (Kraushaar & Novak, 2010): in the case of our students, we commonly observe how they perform their tasks on the computer at the same time as they have their social networks open, and see that they frequently check their mobile phone notifications. Several studies (Kraushaar & Novak, 2010; Kuznekoff & Titsworth, 2013; Purcell et al., 2013; Wammes et al., 2019; Wang et al., 2014; Xu, 2008) indicate that the use of laptops during a lecture distracts the student and those peers who have a direct view of the laptop, thus influencing their learning, since students engage in off-topic activities. What is more, Sana et al. (2013) states that the use of several types of electronic media among university students is the principal cause of low academic performance. Furthermore, a study with 5000 students indicated that 38% answered that they had to check their electronic devices every 10 minutes when they were studying (Zhang, 2015). In Jacobsen & Forste (2010), the authors found that students could maintain their academic activities for only a short amount of time (around 6 minutes), and after that time they needed to move to a technological distraction. We consider that these data are a cause for concern and that we as educators should attempt to do something to remedy this situation.

In Kessler (2011), the author recommend exploring other influential variables as regards academic distractions (other than technology). In our work, we would like to contribute by also studying the interruptions provoked by our own thoughts and by the people around us. An example of the second type of distraction is when a student is working and another student interrupts him/her to ask for help, or something else. An example of the first kind of interruption, meanwhile, occurs when we are focusing on a topic, but a negative or recurrent thought, which is neither important nor relevant at that particular moment, appears and distracts us from our focus. In Was et al. (2019), the authors studied the impact of mind wandering as a factor that decreases learning. The same authors claimed that mindfulness training reduces students' mind wandering and improves their performance, as is also described in Rosen et al. (2013).

Literature contains numerous studies that corroborate the effectiveness of practising *mindfulness* in multiple areas. Among its main achievements, we find: stress reduction, improvement in attention, a reduction in anxiety levels, the treatment of depression, emotional enhancement, better motivation and improved well-being (Feng et al., 2019).

The integration of concentration and *mindfulness* techniques in academic environments (Altmann et al., 2014; Damico & Whitney, 2017; Pidgeon & Keye, 2014; Yildirim & Dark, 2018), technological contexts (Elliott-Dorans, 2018; Fish et al., 2016; Salehzadeh Niksirat et al., 2017; Sana et al., 2013; Waite et al., 2018; Zhu et al., 2017) or those simply prone to interruptions (Elliott-Dorans, 2018; Kuznekoff & Titsworth, 2013; Lyddy & Good, 2017) is also a topic that many authors have studied. Studies such as Altmann et al. (2014), Purcell et al. (2013), Wammes et al. (2019) and Xu (2008) have shown beneficial effects after *mindfulness* has been practised in academic environments. However, in a recent study (Mrazek et al., 2013) that concerned the evaluation of the performance of *Software Engineering* students trained in *mindfulness* when compared to the performance of a control group, the results obtained did not provide sufficiently significant evidence of an improvement. These findings provide a gateway to discussion on the effectiveness and study of this field.

Another technique that is useful to control interruptions is time management, this term alluded to the ability to plan study time and tasks (Effeney et al., 2013). As it is mentioned by Liberus et al. (2019), university students often state to have problems managing the time required to perform their study demands. The usage of time management techniques decreases stress and helps people to gain efficiency and a level of satisfaction in their life and jobs (Hanley et al., 2015). Many studies (Khan, 2015; Liborius et al., 2019; Nadinloyi et al., 2013; Sana et al., 2013; Waite et al., 2018; Zhang, 2015) support the direct relationship between students' correct organization of time and concentration, and their academic achievements. As a specific example of this, in Macan et al. (1990), the authors demonstrated that students who used time management had a better academic performance and experienced less stress than those who did not do so.

# PILOT STUDY

In **Phase 4**, a workshop was organized, whose main goal was **improving students' attention and concentration in lectures and when they are studying.**

In order to attain that goal, students need to know techniques that will help them to be aware of their interruptions, their consequences and how to manage them and to keep their attention on a topic. The techniques that we decided include in the workshop were, therefore:

- **Awareness**: The first thing was to encourage students to detect what interruptions they have and what their sources are. In order to help them with this reflection, the students had to fill in a questionnaire in which they were required to indicate the number of interruptions and choose among the different sources of those interruptions. This would, therefore, make the students aware of them, as being aware of a problem is the first step towards solving it. Waite et al. (2018) indicated that educators need to encourage students' self-regulation of laptop multitasking, and in Bellhäuser et al. (2016) the authors propose a web-based training to foster self-regulated learning. Our goal with this technique is to enable students to discover the importance of controlling their interruptions and learn self-regulation strategies since, as stated in Khan (2015), students are required to self-regulate their attention if they are to obtain good learning results.
- **Concentration techniques**: means or actions intended to avoid unwanted interruptions, thus increasing performance in the development of the planned tasks. For example, the practice of concentration and relaxation techniques based on *body scanning* (Macan et al., 1990) was taught; the students were even encouraged to appreciate certain details that they would not otherwise be aware of in their daily life, in order to attempt to focus on the *present moment* for as long as possible. All these ideas are associated with mindfulness. However, we were of the opinion that the students might be more motivated if we called them "concentration techniques".
- **Thought management**: techniques aimed at identifying the interruptions associated with our thoughts and attempting to rid our mind of those negative and simply useless thoughts. For instance, we provided them with information regarding the distinction between different types of interruptions (internal and external) and thoughts (positive/negative, useful/useless). The goal was to train the students to maintain positive and useful thoughts, but to attempt to ignore negative and useless ones.
- **Time management**: proposed planning techniques and time schematization that allow students to organize their tasks according to their schedules. Time management techniques such as the *Pomodoro* method were also introduced.

After performing the pilot study, we obtained several **lessons**:

- **The workshop was too tiring:** When the pilot study had finished, some students were asked for their opinion about their experience. All of them agreed that four-hour two days sessions were too tiring; moreover, if a student was not able to attend a session one day, s/he missed half of the workshop. This may have been one reason why many students gave up the experience.
- **The follow-up session by Skype was not very useful for the students:** The students had to attend some follow-up sessions, thus enabling the psychologist (via Skype) who taught the course or other teachers to motivate the students to practice the technique learnt during the workshop. The students explained that they preferred the face-to-face meeting with the teachers, as they felt more comfortable and familiar with that format. We consequently decided to eliminate this follow-up session on Skype, opting to increase the number of face-to-face follow-up sessions. This was done in an effort to give and receive continued feedback to and from the students, and to encourage them to practise what they had learnt in the workshop. Moreover, and in an attempt to reduce the drop-out rate, we decided to modify the follow-up sessions so that they would be carried out in pairs; this method could motivate the students, thus inspiring them to share their progress and encourage each other through mutual coaching.
- **Some time management techniques could be explained in a more flexible way**: The students complained that some of the time management techniques taught, such as *Pomodoro,* were not fully adaptable to the tasks performed by computer science students, the duration of which tends to be too long and difficult to be divided into 25-minute periods. The conclusion was, therefore, reached that time management techniques should be more flexible and adaptable to the student's tasks, always respecting rest periods, and avoiding long periods of study during which concentration cannot be maintained.
- **Post-tests did not reflect what interruptions students had been able to control or eliminate.** The questionnaires should, therefore, be revised and improved in order to obtain that particular information.
- **It seemed that the workshop helped students both to reduce their interruptions and to manage their thoughts in a great number of respects**. Not all the lessons learnt were negative; the results seem to indicate that the experience had a positive effect on the students. This encouraged us to perform the following study and to consider that this type of workshop should be carried out in the first year of the degree, in which the drop-out rate and the tendency to fail are high. The most suitable period of time in which to conduct the workshop would be in the first weeks of the second term, when the students do not yet have such a high workload but have already become familiar with the university and are, to some extent, aware that the study methods are different from those of their high schools. Learning the different techniques taught in the workshop from the first year of their degree could help them to modify their study habits throughout the remainder of their academic trajectory.

# EMPIRICAL STUDY

Having considered all the lessons learnt from the pilot study, we designed an empirical study, which is described in this section.

## SUBJECTS

A total of 14 first-year students enrolled on the subject *Programming II* on the Computer Science Degree participated in this empirical study. Furthermore, the psychologist who had taught the pilot workshop came back to teach the improved workshop.

## MATERIAL

The materials used consisted of:
- A pre-test similar to the previous one (see *Appendix A*).
- A modified **post-test,** refined in order to measure not only the number of interruptions, but also how many were handled by the students (see *Appendix C*).
- A modified **follow-up test** to evaluate the progress of those students who had taken part in the workshop (see *Appendix B*). The goal of these sessions was to detect whether the students had practised the techniques taught during the workshop and to discover whether the students had detected any changes in their behaviour.
- A set of **slides** used for the presentation and development of the workshop.
- A series of **audio guides**, provided by the psychologist, with which to carry out concentration practice.

## PROCEDURE

A pre-test was distributed to first-year students enrolled in the workshop. We decided to carry out comparisons with the data obtained in the pre and post-test. This comparison would provide us with conclusions about the effectiveness of the workshop.

In this study, the workshop took place over a total of three days in two-hour sessions, a format that guaranteed that if any of the students enrolled could not attend on one of the days, they would not miss half of the workshop, as had occurred in the pilot study. This format would, in turn, make these sessions more bearable and dynamic than the four-hour sessions that took place in the pilot workshop.

Another structural change was the introduction of three follow-up sessions, each of which was performed on the dates indicated in *Figure 6*. This was done to motivate the students to undertake the tasks they had learnt, and to ensure a better monitoring of their progress. After these sessions, each of the pairs that carried them out was simultaneously entrusted with tasks that would help them to reduce bad habits or eliminate sources of interruption, using what they had learnt in the workshop. A schematic summary of the different sessions is included in *Figure 6*.

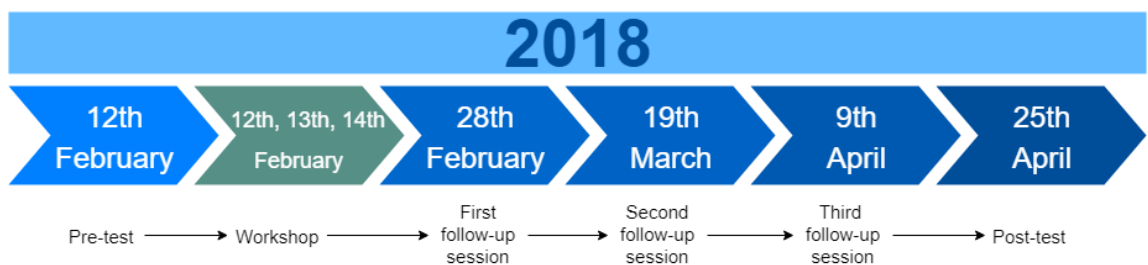

**Fig. 6. Study schedule.**

## VARIABLES AND HYPOTHESES

In order to evaluate our research question: whether *"a workshop in time, attention and thought management helps students to manage their interruptions, improve their level of concentration and also improve their time management"*, we took the training (the workshop) of the students in this topic as an independent variable of our study. Furthermore, the dependent variables were the participants' levels of **time organization**, **thought management** and **concentration**, treated as ordinal values.

The topics under consideration, as well as the stated null and alternative hypotheses of this study are therefore as follows:

- **Time organization** ($H_{01}$):
  - $H_{01}$: *the training did not help the students to improve their time organization.*
  - $H_{11}$: *training in time, attention and thought management improves students' time organization.*

- **Thought management** ($H_{02}$):
  - $H_{02}$: *the training did not help the students to improve their thought management.*
  - $H_{12}$: *training in time, attention and thought management improves students' thought management.*

- **Concentration levels** ($H_{03}$):
  - $H_{03}$: *the training did not improve the students' levels of concentration.*
  - $H_{13}$: *training in time, attention and thought management improves students' levels of concentration.*

The measures of the dependent variables were collected by means of the questionnaires carried out before and after the workshop, and the answers in the pre and post-tests were compared. In order to discover whether the students had improved their skills about the topics under consideration, the following questions were analysed and compared in the pre and post-test:

- Time Organization: **Q14**
- Thought management: **Q15, Q16**
- Concentration levels: **Q17**

Furthermore, it could be considered that when students concentrate on a certain task, less attention is paid to interruptions. On this basis we also assessed whether students felt they had experienced fewer interruptions during a one-hour lecture or in a one-hour study session after participating in the workshop. The questions **Q2, Q4, Q5, Q6, Q7, Q9** were therefore also compared before and after the workshop.

## RESULTS

When comparing the number of students using time management techniques before and after the workshop (see *Table 1*) we detected a significant enhancement: while before the workshop a 43% of the students affirmed that they used some kind of time management technique, after the completion of the workshop 93% of the students stated that they used time management techniques. *Table 1* describes the techniques used, as well as their frequency. These data indicate that the experience was useful in encouraging students to organise their time.

We also tried to study whether the students introduced any thought-management activities into their routines. *Table 2* shows that 50% of the students began to use thought management techniques. The techniques that the students used are described in *Table 2*. The data therefore indicate a significant improvement after attending the workshop.

**Table 1.** Time-management techniques used before and after the workshop.

| Pre | Number | % | Post | Number | % |
|---|---|---|---|---|---|
| Pomodoro | 4 | 28.57 | Pomodoro | 10 | 71.43 |
| *To-do* list | 1 | 7.14 | *To-do* list | 2 | 14.29 |
| Time schedule | 1 | 7.14 | Time schedule | 1 | 7.14 |
| None | 8 | 57.14 | None | 1 | 7.14 |
| | 14 | | | 14 | |

**Table 2.** Thought-management techniques used before and after the workshop.

| Pre | Number | % | Post | Number | % |
|---|---|---|---|---|---|
| Music | 2 | 14.29 | Meditation | 5 | 35.71 |
| None | 12 | 85.71 | Breathing | 3 | 21.42 |
| | 14 | | Music | 1 | 7.14 |
| | | | None | 5 | 35.71 |
| | | | | 14 | |

We shall now focus on the levels of concentration. On comparing *Table 3* and *Table 4* as regards attention span, it will be noted that there is a significant improvement after participating in the workshop, since after doing so, most of the students considered that they had a *Good* level of attention (78.6%). However, in the pre-test, 50% stated that they had a *Normal* attention span, while 14.3% stated that it was *Bad*. Nobody considered themselves to have a *Very good* attention span but, after attending the workshop, at least one student considered his/her attention span as *Very good*.

**Table 3**. Descriptive statistics.
**Level of concentration pre-test**

| | | Frequency | Percentage | Cumulative percentage |
|---|---|---|---|---|
| Levels of attention | Good | 5 | 35,7 | 35,7 |
| | Normal | 7 | 50,0 | 85,7 |
| | Bad | 2 | 14,3 | 100,0 |
| | Total | 14 | 100,0 | |

**Table 4**. Descriptive statistics.
**Level of concentration post-test**

| | | Frequency | Percentage | Cumulative percentage |
|---|---|---|---|---|
| Levels of attention | Very good | 1 | 7.1 | 7.1 |
| | Good | 11 | 78.6 | 85.7 |
| | Normal | 1 | 7.1 | 92.9 |
| | Bad | 1 | 7.1 | 100.0 |
| | Total | 14 | 100,0 | |

After that, the changes in the number of interruptions were analysed. As shown in *Table 5* and *Table 6*, the number of interruptions decreased in both lecture and individual study hours after the completion of the workshop.

**Table 5**. Descriptive statistics.

| Number of interruptions in one hour study pre-test | | Frequency | Percentage | Cumulative percentage |
|---|---|---|---|---|
| Interruptions during one-hour study | 0..5 | 6 | 42.9 | 42.9 |
| | 6..11 | 5 | 35.7 | 78.6 |
| | 12..17 | 2 | 14.3 | 92.9 |
| | 18..22 | 1 | 7.1 | 100.0 |
| | Total | 14 | 100.0 | |

**Table 6**. Descriptive statistics.

| Number of interruptions in one hour study post-test | | Frequency | Percentage | Cumulative percentage |
|---|---|---|---|---|
| Interruptions during one-hour study | 0..5 | 8 | 57.1 | 57,1 |
| | 6..11 | 4 | 28.6 | 85.7 |
| | 12..17 | 2 | 14.3 | 100.0 |
| | Total | 14 | 100.0 | |

As will be noted when comparing *Table 5* and *Table 6*, the percentage of students that reduced the number of interuptions during their study time between 0 and 5 increased by 14% after participating in the workshop and its follow up. It will also be noted that in the post-test, the majority of the students (85.7%) had between 0 and 11 interruptions; no students had more than 17 interruptions, as had occurred in the pre-test data.

With regard to the interruptions during lecture (see *Table 7* and *Table 8*), a decreasing trend was also noted after the workshop. In particular, it is possible to observe an increase of 35% for those students who now had between 0 and 11 interruptions; none of the students had more than 17 interruptions after participating in the workshop. However, before taking part, 42.8% of the students had indicated that they had more than 18 interruptions during a one hour lecture.

**Table 7**. Descriptive statistics.

| Number of interruptions in one- hour lecture post-test | | Frequency | Percentage | Cumulative percentage |
|---|---|---|---|---|
| Interruptions during one-hour lecture | 0..5 | 4 | 28.6 | 28.6 |
| | 6..11 | 3 | 21.4 | 50.0 |
| | 12..17 | 1 | 7.1 | 57.1 |
| | 18..22 | 3 | 21.4 | 78.6 |
| | 23..28 | 3 | 21.4 | 100.0 |
| | Total | 14 | 100.0 | |

**Table 8**. Descriptive statistics.

| Number of interruptions in one-hour lecture post-test | | Frequency | Percentage | Cumulative percentage |
|---|---|---|---|---|
| Interruptions | 0..5 | 5 | 35.7 | 35.7 |
| during one-hour lecture | 6..11 | 7 | 50.0 | 85.7 |
| | 12..17 | 2 | 14.3 | 100.0 |
| | Total | 14 | 100.0 | |

In order to discover whether these data were statistically significant, we analysed whether the distributions of the populations for each question were normal, and we observed that neither the number of interruptions during lecture and study nor the levels of attention presented a normal distribution; this led us to make use of the Kruskal-Wallis test to address our hypothesis. This test is one of the most suitable methods with which to compare populations whose distributions are not normal. The results obtained for each question are summarized in *Table 9* and *Table 10*.

**Table 9**. **Ranges for each variable.**

| | | N | Average range |
|---|---|---|---|
| **Q2.** Interruptions during study | RT-pre | 14 | 15.64 |
| | RT-post | 14 | 13.36 |
| | Total | 28 | |
| **Q4.** Interruptions during lecture | RT-pre | 14 | 16.96 |
| | RT-post | 14 | 12.04 |
| | Total | 28 | |
| **Q17.** Levels of attention | RT-pre | 14 | 18.00 |
| | RT-post | 14 | 11.00 |
| | Total | 28 | |

**Table 10**. **Kruskal-Wallis test results**

| | Q3 | Q4 | Q17 |
|---|---|---|---|
| Chi-square | .643 | 2.735 | 6.418 |
| lg | 1 | 1 | 1 |
| Asymptotic significance | .422 | .098 | .011 |

According to the results of the Kruskal-Wallis test, the only really statistically significant improvement was made to the students' attention levels, while the number of interruptions improved, but not statistically significantly.

- This information makes it possible to confirm $H_{13}$: *the training improved the students' levels of concentration.*

Having analysed the quantitative data, we shall now explain the results obtained from the questionnaire that the students filled in during the last *follow-up* session, to which 12 students responded (questionnaire shown in *Appendix B*).

The objective of the question Q1 was to discover whether the students had performed the meditation activities, discovering that most of students had sometimes applied them.

Related to the previous one, for Q2 students mainly stated that they can concentrate better when they begin to study, with some of them indicating that they can study for longer or they have more willpower to study. The person who responded "nothing" in Q2 was the same person who had stated that s/he had never meditated before studying, signifying that the answer was coherent. It was also interesting to see that nobody chose the options "I need the same amount of time as before" or "I need more time to study". It would therefore appear that meditating before studying led to good results, and that the students felt that they concentrated better and took greater advantage of their time. Moreover, some of them were more motivated to study and spent longer doing so.

Q3 concerned the fact that the students were grouped in pairs, so that they could motivate each other to perform the different practices learnt in the workshop. Remarkable results of the question were that the most selected option was: "It was very positive, and s/he encouraged me", followed by the option "Working with someone else was no different to working on my own".

About Q4, students had to provide their opinions about the role of the tutor. In this case 10 students indicated that it was very positive, and nobody marked the option "Attending the follow up sessions with the tutor was a waste of time".

The Q5 was related to the number of times that they listened to the audio provided by the psychologist, which contained guided meditation and visualisation. The answers with more frecuency were "between 1 and 2 days" and "between 3 and 4 days". Only one student chose the option "Every day", and one chose the option "never" (the one who said that s/he never meditated before studying).

Q6 was concerned with the time of day that was best for them to listen to the audio, discovering that majority of students consider the best time was before going to sleep, followed by "before studying". Only one student opted for "before going to the university".

Q7 is stated in order to know whether the time when they preferred to listen to the audio coincided with the time when they actually did it, being the preferred options "at night" and "before studying". The option "in the evening" was chosen 2 times and it is curious that nobody marked the option "in the morning".

The frequency that students apply any kind of time-planning method (and which of them) is find out by Q8. In this case 6 students coincided as regards stating that they used a *To-Do list*, and another 6 students indicated that they used the *Pomodoro* technique. However, the frequency with which they used them varied to a greater extent: 4 students chose option "*Between 1 or 2 days*", 4 students chose the option "*Between 3 and 4 days*", 3 students said "*Every day*", and only 1 marked "*Between 5 and 6 days*".

Regarding at what extent students had over their thoughts (Q9), mainly students marked "*Every day*", followed by "*Between 5 and 6 days*" and "*Between 3 and 4 days*". Only 1 student chose the option "*Between 1 or 2 days*". It is interesting to note that nobody chose the option "*Not at all*".

The last question (Q10) was included in an effort to evaluate in what sense students considered that the workshop had changed their lives. The most frequent answer was that students had noted an improvement when they woke up in the morning (7), followed by there was no difference (5). Some of them may have slept better because of the audios. Besides, 9 students affirmed that their interruptions and thought management had improved, while 2 of them said that they were no different and 1 stated that they were worse (the one who neither meditated nor listened to the audios). In general students felt that they had improved with the workshops, and in minor proportion (3) some of them did not notice any differences. Only one student wrote that it depended on the subject. With regards to this last point, 8 students chose the option *"better"* and 5 the option *"no difference"*, which is quite similar to the results obtained for the first answer.

When designing the questionnaire, the Q10 was introduced in order to check whether the students answered randomly or just to please the person in charge of the study. The goal was detecting whether the students chose the option *"better"* just to please us, as we think that no difference should be noted in this situation. The answers were: 3 students chosen the option *"better"* and 9 the option *"no difference"*. It would, therefore, appear that the students' answers were honest.

We can conclude, therefore, that the students generally improved in several respects after attending the workshop and the follow-up sessions. These data serve as valuable feedback for our research and as a sign that some of the participants had made individual improvements to certain aspects of their lives.

## THREATS TO VALIDITY

In this section, we shall explore the threats that may have affected our empirical study. The threats to validity shown by Wohlin et al. (2012) will be employed as a basis on which to focus on only those that we considered to be of particular significance in our study; we will also discuss how we have, as far as possible, attempted to mitigate these threats.

With regard to the **validity of the conclusion**, the following threats were identified:

- **Statistical power**. When the power of a statistical test is low, there is a risk of obtaining erroneous conclusions. In our case, one hypothesis was statistically proved, and the descriptive data indicated that the students improved. In the future, more empirical data with which to reinforce the statistical power will be included.
- **Reliability of measures**. Since we were working with subjective data, one of the main purposes of our study was to collect information with a high degree of confidence and veracity. The tests taken by the students were conducted anonymously, without needing to know students' true identity. Thus, although this threat was present in our study due to the presence of subjective variables, we attempted to overcome it in the best possible way.

Furthermore, in the case of the **internal validity** of our experiment, we found the following threats:

- **Testing.** Sometimes when the subjects know that they are being evaluated they can improve their behaviour; in our case we attempted to avoid this effect by using pseudonyms and not obtaining any reward.
- **Instrumentation.** Although this threat was present at the beginning of our experiment, the execution of the pilot study allowed us to refine the format of the tests and improve them on the basis of the review carried out both by their authors and by the psychologist who collaborated in the experiment. That signifies that this threat to our empirical study was very slight.
- **Selection.** It is true that the students who took the workshop may have collaborated with greater motivation than the rest of participants in the experiment. Despite this, as the test questionnaires were conducted with pseudonyms, and the post-test was carried out once the students had been assessed in all their subjects, we believed that the students would not seek to please in a conscious way, as they were aware that there was no advantage in doing so.
- **Mortality.** One of the main problems we had to deal with was that of the people who collaborated in the experiment dropping out, especially in the pilot study (12 of the 16 students who attended the workshop); in the empirical study we added one more face-to-face follow-up session, and that seems to avoid the mortality effect.

In terms of **construction validity**, we identified the following threats:

- **Mono-method bias**. The evaluation of subjective and intrinsic variables for the subjects may influence the bias of the results. Although it is especially difficult to achieve completely objective results in a study of this nature, in the quest to achieve impartial responses, an attempt was made to objectify the questions in the tests as much as possible, and whenever possible.
- **Interaction of testing and treatment**. Although the subjects' training in time, attention and thought management could, in theory, have influenced the results and their awareness of the evaluation activities, no indication was made to any of the students who received this training that a post-test would be carried out.

- **Hypothesis guessing**. As we have seen, the anonymity factor and the lack of rewards associated with test results mitigated this risk.

One thing that can be observed upon studying literature (Feng et al., 2019; Mrazek et al., 2013; Xu, 2008) is that the generalization of the results obtained after carrying out this kind of empirical studies in any type of environment must be treated with caution (there are many factors to consider and generalizations are sensitive). This kind of risk is corroborated by Pidgeon & Keye (2014), signifying that the results obtained from our study are not an exception: the fact that our results have been favourable represents an improvement in this field, further work is required to corroborate the effectiveness of similar training workshops.

# CONCLUSIONS, CONTRIBUTIONS TO THE FIELD AND FUTURE WORK

In this paper, we have tackled how university students could improve their time and attention management by means of suitable learning methods, particularly to reduce the number of interruptions that they have to confront. The effects of interruptions have been widely dealt with in related literature. However, although it has been demonstrated that there is a clear awareness of this problem in the academic world, there is a lack of proposals related to the learning process. In this work, we have attempted to find a possible solution, which is based on providing students with training by means of a workshop focused on time, thought and attention management, with the aim of helping them to better manage their different types of distractions.

As a result of the study presented, the first conclusion that we would like to highlight is the high number of interruptions that students experience in lectures and during their study periods. We are aware that having to study and work with a computer implies that students are closely related to ICT and are prone to interruptions. We can, therefore, state that students in higher education are potentially vulnerable to the main sources of disruption associated with technology. As literature indicates, these interruptions have detrimental effects on our attention and task performance, and are phenomena to which our students are exposed.

As suggested by Liborius et al. (2019) teachers should become involved in order to improve their students' performance. The authors are of the opinion that there is a need for training for students to foster their planning, effort regulations, avoidance of procrastinations etc. We think that a first step is informing the students about: (i) the number of interruptions they are exposed and (ii) these detrimental effects. This will allow teachers to help students be aware of the risks, helping them to change their behaviour. Since, the sooner students are aware of this situation the better, the first academic year could be the most suitable time for that.

The second step is that of providing a workshop, since this seems to be a good technique by which to improve students' concentration. After taking part in the workshop, students are able to do more in less time and can, therefore, make the most of their time in their lectures and study periods. Moreover, they learn how to manage their interruptions and ignore them when they are busy.

Our contribution to the field is, therefore, to propose a method with which to mitigate students' interruptions and train them to manage the different interruptions and to use time management techniques to be more focused and concentrated. We wished to share the results obtained in this study in order to encourage other universities to perform similar activities that would educate students to ignore social networks and other external and internal interruptions whenever possible, which is currently a challenge, as literature explains.

To conclude, we can state that the results obtained are encouraging, and they have motivated us to continue giving this workshop in future academic years, obtain more data and to enable us to analyse its effects. One thing that it would be interesting to see is whether the students who took part in this study obtained better marks than those who did not.

# APPENDIX A. FIRST QUESTIONNAIRE

Gender:      Female ☐   Male ☐   Other ☐

**Q1.** Approximately how much time do you study each day?

        a) 0..30'   b) 31..60'   c) 61..90'   d) 91..120'   e) 121..150'   f) 151..180'   g) 181..210'

**Q2.** Approximately how many interruptions do you experience when studying for one hour (at home, in the library, etc.)?

        a) 0..5   b) 6..11   c) 12..17   d) 18..22   e) 23..28   f) 29..32   g) more than 33

**Q3.** Approximately how many interruptions do you experience during **one hour of lecture**?

        a) 0..5   b) 6..11   c) 12..17   d) 18..22   e) 23..28   f) 29..32   g) more than 33

**Q4.** What are the **external** sources of interruptions **during one hour of study** (at home or where you normally study)?

E.g.: Facebook, WhatsApp, people, calls, other social networks.

Indicate which: _______________________________________________________

**Q5.** What are the **external** sources of interruptions **during one hour of lecture**?

E.g.: Facebook, WhatsApp, people, calls, other social networks.

Indicate which: _______________________________________________________

**Q6.** What are the **internal** sources of interruptions **during one hour of study** (at home or where you normally study)?

E.g.: eating, talking to someone, smoking, going to the bathroom, thoughts that distract you, etc.

Indicate which: _______________________________________________________

**Q7.** What are the **internal** sources of interruptions **during one hour of lecture**?

E.g.: eating, talking to someone, smoking, going to the bathroom, thoughts that distract you, etc.

Indicate which: _______________________________________________________

**Q8.** Do you use any time-management techniques?

**Q9.** Do you use any relaxation techniques?

**Q10.** Do you use any thought-management techniques?

**Q11.** How would you define your attention span?

        a) Very good   b) Good   c) Normal   d) Bad   e) Very bad

# APPENDIX B. EMPIRICAL STUDY FOLLOW-UP TEST

**Q1**. Have you meditated before studying? Always/Sometimes/Never.

**Q2**. What have you noticed? (**choose all the options you wish**)

- Nothing.
- I can concentrate better when I begin to study.
- I can study for longer.
- I get tired more quickly.
- I pay more attention to my thoughts.
- I have more willpower to study.
- I need more time to study.
- I need the same amount of time as before.
- I do more work in less time.

**Q3**. What is your opinion of your partner's monitoring of you?

- Very positive and s/he encouraged me.
- Working with someone else was no different to working on my own.
- I think that working with a partner was more distracting than encouraging.

**Q4**. What is your opinion of your tutor's monitoring of you?

- Very positive and s/he encouraged me.
- It was no different to not having it
- Attending the follow up sessions with the tutor was a waste of time

**Q5**. How often have you managed to listen to the audio (On average)

- Everyday.
- Between 5 and 6 days.
- Between 3 and 4 days.
- Between 1 and 2 days.
- Not at all.

**Q6**. What is the best time of the day for you to listen to the audio? Response:

**Q7**. When do you usually listen to the audio? (Choose all the options that you wish and indicate the duration of the audio at the side

- In the morning, duration:
- Before studying, duration:
- In the afternoon, duration:
- At night, duration:
- Others:______________, duration: ________

**Q8**. Have you used any kinds of time-planning methods? What techniques have you used?

How often?

- Everyday.

- Between 5 and 6 days.

- Between 3 and 4 days.

- Between 1 and 2 days.

- Not at all.

**Q9**. How often are you conscious of the types of thoughts going through your head?

- Every day

- Between 5 and 6 days.

- Between 3 and 4 days.

- Between 1 and 2 days.

- Not at all.

**Q10**. With regard to your management of interruptions and thoughts, please indicate whether the following facets have been modified in your case:

- When I get up in the morning…worse/better/no difference

- When I study…worse/better/no difference

- When I pay attention in lecture…worse/better/no difference.

- When I enjoy my free time…worse/better/no difference.

- When I sleep…worse/better/no difference.

# APPENDIX C. EMPIRICAL STUDY POST-TEST

Gender:   Female ☐   Male ☐   Other ☐

**Q1**. Approximately how much time do you study each day?

      a) 0..30'   b) 31..60'   c) 61..90'   d) 91..120'   e) 121..150'   f) 151..180'   g)181..210'

**Q2**. Approximately how many interruptions do you experience when studying for one hour (at home, in the library, etc.)?

      a) 0..5   b) 6..11   c) 12..17   d) 18..22   e) 23..28   f) 29..32   g) more than 33

**Q3**. Has your conduct as regards this type of interruptions altered after doing this course? How do you now respond to them?

**Q4**. Approximately how many interruptions do you experience during **one hour of lecture**?

      a) 0..5   b) 6..11   c) 12..17   d) 18..22   e) 23..28   f) 29..32   g) more than 33

**Q5**. Has your conduct as regards this type of interruptions altered after doing this course? How do you now respond to them?

**Q6**. What are the **external** sources of interruptions **during one hour of study** (at home or where you normally study)?

E.g.: Facebook, WhatsApp, people, calls, other social networks.

Indicate which: ______________________________________________________

**Q7**. Which of these **external sources** have you managed to control/eliminate, and which have you not?

I have managed to control/eliminate:

I have not managed to control/eliminate:

Because:

**Q8**. What are the **external** sources of interruptions **during one hour of lecture**?

E.g.: Facebook, WhatsApp, people, calls, other social networks.

Indicate which: ______________________________________________________

**Q9**. Which of these **external sources** have you managed to control/eliminate, and which have you not?

I have managed to control/eliminate:

I have not managed to control/eliminate:

Because:

**Q10**. What are the **internal** sources of interruptions **during one hour of study** (at home or where you normally study)?

E.g.: eating, talking to someone, smoking, going to the bathroom, thoughts that distract you, etc.

Indicate which: ______________________________________________________

**Q11**. Which of these **internal sources** have you managed to control/eliminate, and which have you not?

I have managed to control/eliminate:

I have not managed to control/eliminate:

Because:

**Q12**. What are the **internal** sources of interruptions **during one hour of lecture**?

E.g.: eating, talking to someone, smoking, going to the bathroom, thoughts that distract you, etc.

Indicate which: _________________________________________________________

**Q13**. Which of these **internal sources** have you managed to control/eliminate, and which have you not?

I have managed to control/eliminate:

I have not managed to control/eliminate:

Because:

**Q14**. Do you use any time-management techniques?

**Q15**. Do you use any relaxation techniques?

**Q16**. Do you use any thought-management techniques?

**Q17**. How would you define your attention span?

    a) Very good    b) Good    c) Normal    d) Bad   e) Very bad

**Q18**. Has this course fulfilled your expectations? Indicate what has and what has not.